\documentstyle[manuscript,aps]{revtex}
\begin{document}
\draft
\begin{center}
{\LARGE Invariant Correlational Entropy and Complexity of Quantum
States}\\[0.5cm]

{\bf Valentin V. Sokolov$^{1}$, B. Alex Brown$^{2}$ and Vladimir
Zelevinsky$^{1,2}$}\\[0.2cm]

{\sl $^{1}$Budker Institute of Nuclear Physics, 630090 Novosibirsk, Russia\\
$^{2}$Department of Physics and Astronomy and\\
National Superconducting Cyclotron Laboratory,\\
Michigan State University, East Lansing, MI 48824-1321, USA}
\end{center}

\begin{abstract}

We define correlational (von Neumann) entropy for an individual quantum
state of a system whose time-independent hamiltonian contains random
parameters and is treated as a member of a statistical ensemble. This
entropy is representation independent and can be calculated as a trace
functional of the density matrix which describes the system in its
interaction with the noise source. We analyze perturbation theory in order
to show the evolution from the pure state to the mixed one. Exactly solvable
examples illustrate the use of correlational entropy as a measure of the
degree of complexity in comparison with other available suggestions such as
basis-dependent information entropy. It is shown in particular that a
harmonic oscillator in a uniform field of random strength comes to a
quasithermal equilibrium; we discuss the relation between effective
temperature and canonical equilibrium temperature.  The notion of
correlational entropy is applied to a realistic numerical caculation in the
framework of the nuclear shell model. In this system, which reveals generic
signatures of quantum chaos, correlational entropy and information entropy
calculated in the mean field basis display similar qualitative behavior.
\end{abstract} \newpage

\section{Introduction}

The concept of entropy is fundamental for many branches of physics and other
sciences dealing with systems which reveal a certain degree of complexity
and disorder. As stressed in the monograph \cite{QE}, ``entropy is not a
single concept but rather a family of notions". This monograph and the
earlier review article \cite{RMP} contain historical information and give
many examples of different approaches to the idea of entropy and numerous
applications.

In relation to quantum theory, the mainstream of development is formed by
four main overlapping lines. They can be referred to as thermodynamical
(Boltzmann - Gibbs) entropy, quantum ensemble (von Neumann) entropy,
information (Shannon) entropy and dynamical (Kolmogorov - Sinai) entropy.
Since the general description of a quantum system, including its interaction
with the environment, time development and relaxation to equilibrium, can be
given in terms of the density matrix \cite{Land,Blum}, the von Neumann
definition seems to be the most fundamental. For a system in an equilibrium
with a heat bath, the density matrix (and, accordingly, von Neumann entropy)
is eqivalent to that in the canonical or grand canonical thermal ensemble.
The evolution of a closed quantum many body (gas-like) system from a random
initial state is shown \cite{Hov,Mark} to lead to the same values of
macroscopic observables as for the thermal equilibrium described by the
microcanonical ensemble which has a clear semiclassical limit as the
equipopulation on the energy surface in phase space. The ensemble entropy
cannot be represented as an expectation value of a dynamic variable expressed
by an operator in Hilbert space.
However, being a trace functional of the density matrix, it is invariant
under unitary basis transformations. For a pure wave function,
as that of a stationary state in an isolated system,
basis-independent von Neumann entropy vanishes. It reflects the mixed
character of the quantum state with incomplete information.

Information entropy, with traditional applications in communication theory,
is expressed in terms of probabilities rather
than amplitudes. Therefore it is representation dependent being different for
different choices of the set of mutually excluding events. In quantum systems,
one can find information entropy of individual eigenstates with respect
to a specific basis. All correlations between the amplitudes of different
components of the wave function are suppressed in this definition. Averaging
information entropy over some ensemble of quantum states, one gets a measure of
average complexity of those states. At this stage, the similarity between
information entropy and von Neumann ensemble entropy can emerge if one can
establish an appropriate correspondence between the ensembles used in the two
approaches and the basis utilized in calculating information entropy
\cite{big}. Thus, for canonical equilibrium thermal ensembles,
the correlations are destroyed by the random interaction with the heat bath
so that the density matrix is diagonal
in the energy representation for the system. In this case, the eigenvalues
of the density matrix give the occupancies of the stationary eigenstates of
the isolated system which could be directly used for constructing
information (= thermodynamical) entropy \cite{Land,Pol}.

The so-called dynamical entropy, extensively studied during the last decade
\cite{Srin,Kos,Pech,AF,Part,Slom}, is essentially information entropy
applied to a random sequence of measurements of quantum observables. Apart
from the intrinsic complexity of the system under study, this construction
reflects special features of the quantum measurement process. This entropy
depends not only on the initial state but, in addition, on the observable
and on the way of performing the measurement. It can be defined so that it
give classical Kolmogorov - Sinai entropy \cite{LL} in the corresponding
limit of fine-grained phase space when the sequence of measurements may be
described with the aid of symbolic dynamics \cite{HT,Lind,Beck,Slom}.

Information entropy used as a tool for quantifying the
degree of complexity of individual quantum states \cite{Yon,Reichl,Izr,MF}
shows delocalization of the wave function in a given basis. However, as a
rule one can find a basis, or a family of bases, which are singled out by
physical arguments specific for each system. The delocalization length in
such a representation manifests complex character of the state and can be
quantitatively related to other signatures of quantum chaos. This
exceptional role is naturally played by the coordinate representation in
billiard-like cases \cite{Reichl} and by the quasienergy basis in the
problems with a periodic perturbation \cite{Izr}. For realistic many-body
systems with strong interaction between constituents, the mean field
represents the exceptional basis where the local correlations and
fluctuations of adjacent stationary states are separated from their regular
evolution along the spectrum \cite{MF}.

As shown in large scale nuclear shell model calculations \cite{entr,big},
the representation dependence of information entropy might be considered in
some respects as an advantage which provides a useful physical measure of
mutual relationship between the eigenbasis of the hamiltonian and the
representation basis. Moreover, chaotic dynamics make different states with
close excitation energy and the same values of exact constants of motion
``look the same" \cite{Perc}, i.e. have similar observable properties.  This
is nothing but a microscopic picture of thermal equilibrium \cite{ann}.
After averaging over a narrow energy window in a high level density region,
information entropy in the mean field basis becomes a smooth function of
excitation energy and carries \cite{temp,big} the same thermodynamic
contents as thermal entropy found for the microcanonical distribution from
the level density.  Being calculated in a random basis, the magnitude of
information entropy of generic states in a complex system is typically on
the level predicted by random matrix theory and does not display any regular
evolution along the spectrum.

The goal of the present paper is to explore the possibility of describing
the degree of complexity of individual quantum states using the von Neumann
definition of entropy and applying an external noise which converts a pure
state into the mixed one. We do not consider the perturbation to be weak;
therefore the resulting mixed state depends explicitly on the noise
properties and gives a description of the ensemble ``system plus noise".
There exists a vast literature discussing the behavior of regular and
chaotic systems under the change of parameters of the hamiltonian, see for
example \cite{pechuk,yuk,sza,simA,sim,alh,gasp,zak,opp,kus,kus1}. Multiple
avoided crossings of the energy terms in a function of parameters reveal
strong mixing and drive the system to the chaotic limit. The analogy of
level dynamics with that of the one-dimensional gas of colliding particles
is very productive for studying the spectral statistics
\cite{Dyson,pechuk,yuk,zak,haake,sok}. Here we assume that the ensemble of
the parameter values is defined by a distribution function and calculate the
density matrix and von Neumann entropy for a given energy term. Using
exactly solvable models, we show essential features of
representation-independent entropy obtained according to this definition,
its similarity to and distinction from information entropy. Even for the
simplest systems as a harmonic oscillator in a random uniform field, the
resulting steady states are far from trivial.  We also give an example of a
realistic numerical calculation for a many-body system of fermions (a
nucleus $^{24}$Mg) which shows that our ensemble entropy (``correlational"
entropy) is a smooth function of excitation energy and therefore may be used
as a measure of the degree of complexity.

\section{Density matrix and correlational entropy}

We consider a quantum system interacting with a surrounding. The interaction
will be parameterized by a set of real  parameters $\lambda$ in the
hamiltonian, $H=H(\lambda)$. The energy spectrum of the system is assumed to
be discrete. The eigenfunctions $|\alpha;\lambda \rangle$ of the system, as
well as its energy levels $E_{\alpha}(\lambda)$, evolve with $\lambda$.  For
a complicated system, the level crossings are avoided so one can
continuously follow these energy terms.

At a fixed value of $\lambda$, one can use any complete orthonormal basis
$|k\rangle$ to study the evolution of the eigenstates in terms of the
amplitudes $C^{\alpha}_{k}(\lambda)$,
\begin{equation}
|\alpha;\lambda\rangle=\sum_{k}C^{\alpha}_{k}(\lambda)|k\rangle. \label{1}
\end{equation}
Instead of the wave function (\ref{1}), one can also use the density matrix
$\rho^{(\alpha)}$ whose elements are
\begin{equation}
\rho^{(\alpha)}_{kk'}(\lambda)=C^{\alpha}_{k}(\lambda)C^{\alpha\ast}_{k'}
(\lambda).                                                  \label{2}
\end{equation}
$\rho^{(\alpha)}$ is a hermitian
matrix in Hilbert space of the system. For a pure state
$|\alpha;\lambda\rangle$, the descriptions in terms of the wave function
(\ref{1}) and the density matrix (\ref{2}) are fully equivalent.  The
obvious properties of the density matrix (\ref{2}) are the normalization
\begin{equation}
{\rm Tr}\,\rho^{(\alpha)}(\lambda)=1                          \label{3}
\end{equation}
and the matrix identity
\begin{equation}
\left(\rho^{(\alpha)}(\lambda)\right)^{2}=\rho^{(\alpha)}(\lambda) \label{4}
\end{equation}
which shows that the eigenvalues of this matrix can be only 0 or 1.
Actually, the density matrix (\ref{2}) is diagonalized in the eigenbasis
$|\alpha';\lambda\rangle$. Only one eigenvalue, for the original state
$\alpha'=\alpha$, is equal to 1 while the rest of the eigenstates belong to
the degenerate subspace with the zero eigenvalues of
$\rho^{(\alpha)}(\lambda)$ so that $\rho^{(\alpha)}(\lambda)$ is the projection
operator onto the state $|\alpha;\lambda\rangle$.  Inversely, these
properties can be taken as a signature of a density matrix describing a pure
state.

Now we assume that the interaction parameters $\lambda$ are random and
have to be considered as members of an ensemble characterized by the
normalized distribution function ${\cal P}(\lambda)$, $\int d\lambda\,{\cal
P}(\lambda)=1$. Then the description in terms of a wave function becomes
impossible, and our system is described by the density matrix (here and
below the overline refers to ensemble averaged quantities)
\begin{equation}
\rho^{(\alpha)}_{kk'}=\overline{C^{\alpha}_{k}C^{\alpha\ast}_{k'}}
=\int d\lambda {\cal P}(\lambda)\rho_{kk'}^{(\alpha)}(\lambda). \label{5}
\end{equation}
This is still a hermitian matrix with trace equal to 1. But, generally, the
operator identity (\ref{4}) is not valid anymore. The eigenvalues
$\rho^{(\alpha)}_{\nu}$ of the matrix (\ref{5}),
\begin{equation}
\rho^{(\alpha)}|\nu)=\rho^{(\alpha)}_{\nu}|\nu),       \label{5a}
\end{equation}
are nonnegative numbers between 0 and 1,
\begin{equation}
(\rho^{(\alpha)}_{\nu})^{2}\leq \rho^{(\alpha)}_{\nu}.   \label{6}
\end{equation}
These eigenvalues can be interpreted as mean occupation numbers of the
eigenstates $|\nu)$ for a system which was brought into the contact with an
external source being originally in the intrinsic state $|\alpha\rangle$.
The exceptional case of a pure wave function is recovered for a fixed
parameter, ${\cal P}(\lambda)=\delta(\lambda-\lambda_{0})$. In notations of
eq. (\ref{5a}) and later on we distinguish the eigenstates $|\nu)$ of the
density matrix from the eigenstates $|\alpha\rangle$ of the hamiltonian by
using the parentheses and the angular brackets, respectively; the dimensions
of both sets are equal.

The statistical distribution of occupancies can be characterized by
von Neumann entropy
\begin{equation}
S^{(\alpha)}=-{\rm Tr}\,\left\{\rho^{(\alpha)}\ln\rho^{(\alpha)}\right\}
=-\sum_{\nu}\rho^{(\alpha)}_{\nu}\ln\rho^{(\alpha)}_{\nu}.       \label{7}
\end{equation}
This entropy, being still attributed to a single original energy term
$|\alpha\rangle$, reflects correlational properties of the system subject to
different levels of noise. Therefore we will call it {\it correlational
entropy} although the definition (\ref{7}) is quite similar to that of
standard thermodynamic entropy in canonical thermal ensembles \cite{RMP}.
In contrast to information entropy of a given complicated state in a fixed
basis $|k\rangle$, which was used in the studies of quantum chaos
\cite{Yon,Reichl,Izr,entr},
\begin{equation}
I^{\alpha}=-\sum_{k}W^{\alpha}_{k}\ln W^{\alpha}_{k}, \quad W_{k}^{\alpha}=
(C^{\alpha}_{k})^{2},                                            \label{7a}
\end{equation}
correlational entropy (\ref{7})
is invariant and does not depend on the original basis
$\{|k\rangle\}$ of simple configurations. Obviously, this is a consequence of
correlations between different components of the eigenfunction which are absent
in the probabilistic definition of eq. (\ref{7a}).

For a pure quantum state $|\alpha;\lambda\rangle$, correlational entropy
vanishes independently of the degree of complexity of the system. Thus,
$S^{\alpha}$ characterizes the intrinsic term $|\alpha\rangle$ as a member
of the statistical ensemble. In general, $S^{\alpha}$ has an order of
magnitude of $\ln \tilde{N}_{\alpha}$ where $\tilde{N}_{\alpha}$ is a number
of the eigenstates $|\nu)$ which have the occupancies
$\rho^{(\alpha)}_{\nu}$ noticeably different from zero. The maximum possible
value of correlational entropy is $\ln N$ where $N$ is the dimension of
Hilbert space. This value would correspond to the ``microcanonical" density
matrix, $\rho^{(\alpha)}_{\nu}={\rm const}=1/N$. Note that the information
entropy \cite{big} has the order of magnitude of $\ln N_{\alpha}$ where
$N_{\alpha}$ is a number of essential components in the stationary wave
function $|\alpha;\lambda\rangle$ expressed in an original basis which was
used in the definition of information entropy. Although formally the maximum
value of information entropy is also $\ln N$, its typical value in the
Gaussian orthogonal ensemble of random real symmetric hamiltonians stays,
due to the fluctuations, on the level of $\ln(0.48 N)$, see
\cite{Izr,Jones,big}. In contrast to that, correlational entropy shows the
degree of mixing, or decoherence, introduced by a given source of noise,
regardless of the relationship between the resulting state and the original
unperturbed basis.

In next two sections we consider simple examples which allow us to obtain
exact solutions and shed some light on main properties and physical meaning
of new entropy.

\section{Perturbation theory}

We start with the case of a narrow noise range $\Delta \lambda$ that is
small as compared to the scale of the parameter values which would lead to a
considerable change of the wave functions. This is the perturbative regime.
We can assume that the distribution function ${\cal P}(\lambda)$ is
concentrated near $\lambda=0$ and find the perturbed wave function which
starts its evolution for $\lambda\neq 0$ from the unperturbed state
$|0\rangle$. The state $|0\rangle$ acquires the admixtures of states $|k\neq
0\rangle$ which are given by standard perturbation theory (we assume the
absence of degeneracy).

With the hamiltonian $H=H_{0}+\lambda V$ where the perturbation $V$ is an
off-diagonal operator in the eigenbasis of $H_{0}$, the density matrix
(\ref{5}) of the state $|0\rangle$ is, up to the second order in $\lambda$,
$$\rho_{lm}=\delta_{l0}\delta_{m0}\left(1-\overline{\lambda^{2}}
\sum_{k\neq 0}
\frac{|V_{k0}|^{2}}{\epsilon_{k}^{2}}\right)-
\overline{\lambda}\left(\delta_{l0}\frac{V^{\ast}_{m0}}{\epsilon_{m}}
+\delta_{m0}\frac{V_{l0}}{\epsilon_{l}}\right)$$
\begin{equation}
+\overline{\lambda^{2}}\left\{\frac{V_{l0}V_{m0}^{\ast}}{\epsilon_{l}
\epsilon_{m}}+\frac{\delta_{l0}(1-\delta_{m0})}{\epsilon_{m}}\sum_{k\neq 0}
\frac{V^{\ast}_{mk}V^{\ast}_{k0}}{\epsilon_{k}}+\frac{\delta_{m0}(1-
\delta_{l0})}{\epsilon_{l}}\sum_{k\neq 0}\frac{V_{lk}V_{k0}}{\epsilon_{k}}
\right\}.                                                 \label{7b}
\end{equation}
Here $\overline{\lambda}$ and $\overline{\lambda^{2}}$ are the mean values
over an ensemble of noise, and the notation $\epsilon_{k}\equiv E_{k}-E_{0}$
is used for the energy denominators.  The density matrix (\ref{7b})
incorporates two effects, the redefinition of the original wave function
$|0\rangle$ due to the perturbation $\lambda V$, and the transition from the
pure state to the density matrix. The first effect is the only one if the
perturbation is fixed while the second effect appears because of the
ensemble distribution of perturbations.

It is easy to see that the first order correction to the density matrix does
not change the purity condition (\ref{4}), $\rho^{2}=\rho$. The decoherence
occurs only in the second order. The diagonalization problem (\ref{5a}) for
the matrix (\ref{7b}) can be solved to give, within a needed accuracy, two
nonvanishing eigenvalues $\rho_{\nu}$ as the roots of the
characteristic equation
\begin{equation}
\rho(1-\rho)-w\overline{(\Delta\lambda)^{2}}=0,         \label{7b1}
\end{equation}
and $N-2$ zero eigenvalues.
As expected, the statistical mixing is driven by the mean square fluctuation
$\overline{(\Delta\lambda)^{2}}=\overline{\lambda^{2}}-
\overline{\lambda}^{2}$ of the noise level. The decoherence rate is
determined by the joint action of all virtual transitions,
\begin{equation}
w=\sum_{k}|v_{k}|^{2}, \quad v_{k}\equiv \frac{V_{k0}}{\epsilon_{k}}.
                                                              \label{7b2}
\end{equation}

As seen from (\ref{7b1}), only one state $|0)$ has the eigenvalue $\rho_{0}$
close to 1,
\begin{equation}
\rho_{0}=1-w\overline{(\Delta\lambda)^{2}}.                    \label{7b3}
\end{equation}
The corresponding eigenfunction $|0)=\sum_{k}\psi^{(0)}_{k}|k\rangle$ has a
large component $\psi^{(0)}_{0}\approx1-(w/2)\overline{\lambda}^{2}$ of the
unperturbed state $|0\rangle$ and small admixtures of other unperturbed
states, $\psi_{k}^{(0)}\approx -\overline{\lambda}v_{k}$, $k\neq 0$; the
corrections are of higher order if the distribution function ${\cal
P}(\lambda)$ is even so that $\overline{\lambda}=0$.

In the approximation (\ref{7b1}), the new state $|1)$ appears with a small
but nonzero occupation factor
\begin{equation}
\rho_{1}=w\overline{(\Delta\lambda)^{2}}.                       \label{7b4}
\end{equation}
The eigenfunction of this state is localized mainly in the subspace
orthogonal to the unperturbed state, $\psi^{(1)}_{k}\approx v_{k}w^{-1/2}$,
$k\neq 0$.  The presence of noise removes the isotropic degeneracy of this
$(N-1)$-dimensional subspace by singling out the direction of the
multidimensional vector of transition amplitudes $\{v_{k}\}$. Finally, the
$N-2$ degenerate states with the zero occupation factors $\rho_{\nu}$ are
orthogonal to the first two states. The high order perturbative corrections
consequentially lift the remaining isotropy populating new combinations of
original states.  The decoherence process can be rather fast due to the
added contributions of many distant admixtures so that perturbation theory
can be valid at a very low noise level only. This ``coherent decoherence"
was discussed in a different context in \cite{bul}.  It is related to the
selection of the most important (rainbow) diagrams in theory of disordered
solids and in random matrix theory.

The perturbed occupancies lead to nonzero entropy (\ref{7})
\begin{equation}
S=-\rho_{0}\ln\rho_{0}-\rho_{1}\ln\rho_{1}\approx W(1-\ln W), \quad
W=w\overline{(\Delta\lambda)^{2}}.                      \label{7b5}
\end{equation}
The singularity at the origin implies the infinite slope, $dS/dW=-\ln W$,
of growing entropy.

\section{Two-level systems}

\subsection{A two-level system with fluctuating level positions}

The simplest nontrivial but analytically solvable problem of a two-level
system illustrates how correlational entropy is related to an interplay
between the off-diagonal level mixing and their random diagonal
displacements induced by the change of the parameter $\lambda$ within a
range $\sim\Lambda$ defined by an ensemble ${\cal P}(\lambda)$.

Here we consider two interacting states with a fluctuating spacing so that
the hamiltonian can be written, with the help of spin matrices, as
\begin{equation}
H=\frac{1}{2}(\epsilon-\lambda)\sigma_{z}+V\sigma_{x}.           \label{8}
\end{equation}
The diagonalization of this hamiltonian at a fixed value of $\lambda$ is
performed by the basis rotation through an angle $\varphi/2$ defined by
\begin{equation}
\sin\varphi=\frac{2V}{\Delta (\lambda)},
\quad \cos\varphi=\frac{\epsilon-\lambda}{\Delta (\lambda)},      \label{9}
\end{equation}
where
\begin{equation}
\Delta (\lambda)=\sqrt{(\epsilon-\lambda)^{2}+4V^{2}}              \label{9a}
\end{equation}
is the level spacing.  The amplitudes of the upper, $(+)$, and of the
lower, $(-)$, state can be written in the basis used in eq. (\ref{8}) as
\begin{equation}
C^{(+)}_{1}=\cos\frac{\varphi}{2}, \quad C^{(+)}_{2}=\sin\frac{\varphi}{2},
\quad C^{(-)}_{1}=-\sin\frac{\varphi}{2}, \quad C^{(-)}_{2}=
\cos\frac{\varphi}{2}.                                         \label{10}
\end{equation}

The instantaneous density matrix (\ref{2}) for each of the two states,
\begin{equation}
\rho^{(\pm)}(\lambda)=\frac{1}{2}\left[1\pm(\sigma_{x}\sin\varphi +
\sigma_{z}\cos\varphi)\right],                             \label{11}
\end{equation}
has the eigenvalues 1 and 0, and the corresponding entropy vanishes. After
averaging over the ensemble of the values of $\lambda$, we come to
\begin{equation}
\rho^{(\pm)}=\frac{1}{2}\left[1\pm(\sigma_{x}s+\sigma_{z}c)\right]  \label{12}
\end{equation}
where $c\equiv\overline{\cos\varphi}$ and $s\equiv \overline{\sin\varphi}$ are
averaged quantities which in generally do not satisfy $s^{2}+c^{2}=1$. This
matrix has two nonzero eigenvalues,
\begin{equation}
\rho_{\nu}=\frac{1}{2}(1+\nu r), \quad r=\sqrt{s^{2}+c^{2}}, \quad
\nu =\pm 1,                                         \label{13}
\end{equation}
which are the same for both original states $(\pm)$.
Therefore both states possess the same correlational entropy
\begin{equation}
S=-\frac{1}{2}\left[(1+r)\ln\frac{1+r}{2}
+(1-r)\ln\frac{1-r}{2}\right].                                   \label{14}
\end{equation}
The exact result depends on the ensemble but, as a function of the averaging
range $\Lambda$, the entropy value (\ref{14}) evolves from $S=0$ at
$\Lambda=0$ to
$S=\ln 2$ for the equipartition of the occupancies at large $\Lambda$ when
$\rho_{\nu}=1/2$.

As an example, we consider the behavior of correlational entropy for the
ensemble with the uniform distribution of $\lambda$ in the interval
$(-\Lambda,\Lambda)$. With the definition (\ref{9a}) we have in this case
\begin{equation}
s=\frac{V}{\Lambda}\ln\frac{\epsilon+\Lambda+\Delta(-\Lambda)}
{\epsilon-\Lambda+\Delta(\Lambda)}, \quad
c=\frac{1}{2\Lambda}\left[\Delta(-\Lambda)-
\Delta(\Lambda)\right],             \label{16}
\end{equation}
The simplest case corresponds to the original degeneracy, $\epsilon=0$, when
\begin{equation}
s=\frac{\tau}{2}\ln\frac{\sqrt{1+\tau^{2}}+1}{\sqrt{1+\tau^{2}}-1},\quad c=0,
\quad \tau=\frac{2V}{\Lambda}.          \label{17}
\end{equation}
The averaged parameter $s$ is small, $s\approx \tau\ln(2/\tau)$, if the
noise level is high and the intrinsic mixing is weak compared to the
typical random level spacing, $\tau\ll 1$.  At strong mixing, $\tau \gg 1$,
$s$ reaches 1.  In this limit the levels are split by the dynamic
interaction and the noise is ineffective in changing the population and
reaching the decoherence.  The entropy value correspondingly evolves from
$\ln 2$ to 0. This behavior is opposite to the evolution of information
entropy of the same states,
\begin{equation}
I=-\frac{1+\cos\varphi}{2}\ln\frac{1+\cos\varphi}{2}-\frac{1-\cos\varphi}{2}
\ln\frac{1-\cos\varphi}{2},                                    \label{18}
\end{equation}
as a function of the mixing strength for a fixed diagonal spacing;
$I\rightarrow 0$ at weak mixing and $I\rightarrow \ln 2$ at strong mixing.
In this example, information entropy measures merely the delocalization of the
eigenstates with respect to the original basis.

In Fig. 1 we show correlational entropy (\ref{14}) as a function of $\tau$
for different values of the ratio $\chi= \epsilon/\Lambda$.  At $\chi<1$
(Figs. 1{\sl a} and 1{\sl b} correspond to $\chi=0.2$ and 0.6, respectively),
the results are qualitatively similar to those at $\epsilon=0$, eqs.
(\ref{14}) and (\ref{17}). With no dynamic mixing, $\tau=0$, entropy is
close to its maximum value, $S\approx \ln 2 -\chi^{2}$ at small $\chi$. As
the relative strength $\tau$ of the dynamic mixing increases, the value of
entropy drops, $S\approx [1+\ln(4\tau)]/(4\tau)$ at large $\tau$.  At
$\chi\geq 1$, the entropy behavior changes since the noise level is not
sufficient for covering the original level spacing, see Figs. 1{\sl c} and
1{\sl d}, for $\chi=1$ and 5, respectively.  In this situation, weak dynamic
mixing increases entropy which starts from zero at $\tau=0$ and reaches its
maximum value, much lower than $\ln 2$, at $\tau=\chi$. With the mixing
strength increasing further, the level repulsion prevails which leads again
to very low entropy.

A complementary picture of Fig. 2 shows correlational entropy (\ref{14}) as
a function of the relative noise strength $\chi^{-1}=\Lambda/\epsilon$ for
four different values of the intrinsic dynamical mixing
$\zeta=2V/\epsilon=\tau/\chi$.  Generally, entropy increases with $\chi^{-1}$
and asymptotically approaches the maximum value $\ln 2$. For weak dynamical
mixing $\zeta$, the growth of correlational entropy is very steep in the
vicinity of the point $\chi=1$ where the accidental level crossing is
possible for the first time, compare the change of behavior at this point in
Fig. 1. The stronger is the dynamical mixing the smoother gets the increase
of entropy. Entropy remains low as long as the intrinsic level repulsion
governed by the parameter $\zeta$ dominates.

\subsection{A two-level system with fluctuating mixing}

We consider the same hamiltonian (\ref{8}) assuming now that the diagonal
elements are fixed (we change the notation $(\epsilon-\lambda)\rightarrow
\epsilon$) but the off-diagonal coupling $V$ fluctuates. For definiteness, we
assume that $V$ is uniformly distributed between $v>0$ and $-v$.
This situation is
simpler than that considered in the previous subsection because the only
relevant parameter is that of perturbation theory, the ratio
\begin{equation}
\kappa=\frac{2v}{\epsilon}                                 \label{19}
\end{equation}
of the level spacing due to the mixing to the unperturbed spacing.

The eigenvalues of the density matrix are still given by eq. (\ref{13}). Due
to symmetry of perturbation, the average value of $\sin\varphi$ vanishes
while the average value of $\cos\varphi$ is
\begin{equation}
c=\frac{1}{\kappa}\ln(\kappa+\sqrt{1+\kappa^{2}}).               \label{20}
\end{equation}
With weak mixing,
$c\approx 1-(\kappa^{2}/6)$, and entropy (\ref{14}) is low,
\begin{equation}
S\approx (\kappa^{2}/12)\ln(\kappa^{2}/12), \quad \kappa\ll 1.       \label{21}
\end{equation}
When the mixing intensity increases, $c$ falls down, and
entropy approaches its limiting value for equipopulated levels,
\begin{equation}
S\approx \ln 2-\frac{c^{2}}{2}\approx \ln 2-\frac{1}{2}
\left(\frac{\ln 2\kappa}{\kappa}\right)^{2}. \quad \kappa\gg 1.   \label{22}
\end{equation}
One can note that here the off-diagonal mixing plays a ``natural" role of noise
which creates entropy, in contrast to the previous example when it has been
stabilizing the system by generating dynamic repulsion as a counterpoise to
random noise; that situation can be considered as a prototype of the
localization in disordered solids when the levels in the wells connected by
tunneling fluctuate.

\subsection{Spin in a random field}

This two-state model is more general in the sense that both the mixing
strength and the level spacing are fluctuating here. On the other
hand, the level spacing in this model does not depend on noise so there is
no level crossings.  The hamiltonian describes a spin 1/2 aligning along a
random field,
\begin{equation}
H=\vec{\sigma}\cdot{\bf n},                                \label{19s}
\end{equation}
where the unit vector ${\bf n}$ has a random direction $(\theta,\varphi)$.
Information entropy (\ref{7a}) of the eigenstates for a fixed field
orientation,
\begin{equation}
I^{+}=I^{-}=-\frac{1}{2}\left\{\ln\frac{1-\cos^{2}\theta}{4}+|\cos\theta|
\ln\frac{1+|\cos\theta|}{1-|\cos\theta|}\right\},            \label{21s}
\end{equation}
does not depend on the azimuthal angle $\varphi$. The density matrices
(\ref{2}) of the eigenstates $|\pm\rangle$ with the eigenenergies $\pm 1$
can be written as
\begin{equation}
\rho^{(\pm)}({\bf n})=\frac{1}{2}(1\pm\vec{\sigma}\cdot{\bf n})  \label{20s}
\end{equation}
and depend, contrary to (\ref{21s}), on both coordinate angles.

After the averaging over the ensemble of angles, we find the eigenvalues
$\rho_{\nu}$ of the density matrix (\ref{5})
\begin{equation}
\rho_{\nu}=\frac{1}{2}(1+\nu r), \quad \nu=\pm 1,              \label{22s}
\end{equation}
for any of the two original states $(\pm)$. Here (compare eq. (\ref{13}))
\begin{equation}
r\equiv\sqrt{{\overline{\bf n}}^2}=\sqrt{c^{2}+s_{+}s_{-}}, \quad
c\equiv\overline{\cos\theta}, \quad s_{\pm}
\equiv\overline{\sin\theta\exp(\pm i\varphi)};\qquad
0\leq r\leq 1.                                                     \label{23}
\end{equation}
This allows one to find correlational entropy in the form (\ref{14}),
\begin{equation}
S=-\frac{1}{2}\left\{\ln\frac{1-r^{2}}{4}+r\ln\frac{1+r}{1-r}\right\}.
                                                              \label{24}
\end{equation}
The expressions (\ref{21s}) and (\ref{24}) have identical structure.
Correlational entropy reduces to information one in the case when the
field direction is uniformly distributed on the surface of a cone with the fixed
polar angle $\theta$. The average density matrix is then diagonal in the
original $z$-representation which is singled out by asimuthal symmetry of
the noise distribution. The two entropies coincide numerically at a
point $\theta=\theta_{0}$ where the effective angle $\theta_0$ is defined
by the condition $\cos\theta_{0}=r$.

\subsection{Random matrix ensembles}

The $2\times 2$ Hamiltonian matrix of an arbitrary two-level system can
be presented as
\begin{equation}
H=\frac{1}{2}\left(u+\vec{\sigma}\cdot{\bf v}\right)    \label{G1}
\end{equation}
in terms of one scalar, $u={\rm Tr} H$, and one vector, ${\bf v}=v{\bf n}$,
parameters. The vector ${\bf v}$
is restricted to the $(x,z)$ plane if time reversal invariance
holds and becomes three-dimensional otherwise. Its magnitude $v$ gives
the level spacing. Both entropies $I$ and $S$ depend only on the unit vector
${\bf n}$ so that formulae (\ref{21s}-\ref{24}) of the
previous subsection remain
valid independently of statistical properties of $u$ and $v$.

If all parameters of a hamiltonian matrix or some part of them can be
random, the hamiltonian belongs to a random matrix ensemble. Gaussian
orthogonal, GOE, and Gaussian unitary, GUE, ensembles of large random
matrices attract special attention for modeling chaotic dynamics in
quantum systems.  In Gaussian ensembles, the hamiltonian matrix elements
calculated in a fixed basis are taken as mutually independent Gaussian
random variables. The idea of complete chaos implies that their distribution
be invariant with respect to the choice of the representation basis
$|k\rangle$. This invariance provides a rigid connection between variances
of all matrix elements. However, it has been discovered that one gains
additional possibility of describing special effects such as dynamical
localization by permitting more flexible conditions for the variances. Some
preferential basis there exists then where the localization takes place.
This basis should be defined in each case in accordance with the concrete
physical situation. Basis dependent information entropy or/and inverse
participation ratio are used to describe the localization, see for example
studies of the banded random matrix ensemble \cite{cas,fyod}. Correlational
entropy seems to present a universal and invariant characteristic of chaotic
dynamics in presence of localization.  We hope to return to this problem
elsewhere.

In Gaussian random matrix ensembles, all eigenstates have the same statistical
distributions so that any of them can represent the generic features.
Here we will only point out the simplest properties of correlational entropy
using the example of two-level Gaussian ensembles. According to
(\ref{23}-\ref{24}), this entropy reaches its maximum value of $\ln 2$ under
the condition of
$\overline {\bf n}= 0$. This is obviously the case for invariant
(orthogonal or unitary) ensembles where the vector {\bf n} is distributed
isotropically. For both the ensembles, mean information entropy,
\begin{equation}
\overline {I_{GOE}} =
\frac{1}{2\pi}\int_0^{2\pi}d\varphi\,I(\varphi) = 2\ln2-1,     \label{IGOE}
\end{equation}
or
\begin{equation}
\overline {I_{GUE}}=\frac{1}{4\pi}\int d\Omega\,I(\theta)=\frac{1}{2},
                                                              \label{IGUE}
\end{equation}
is always lower than $S$. In eqs. (\ref{IGOE}) and (\ref{IGUE}) the angular
coordinates of the unit vector ${\bf n}$ are used, $0\leq \varphi\leq 2\pi$ and
$0\leq\theta\leq \pi$.
The instantaneous quantity $I({\bf n})$ passes the maximum
value of $\ln 2$ at the certain orientations of the vector ${\bf n}$,
as at the point $\theta=\pi/2$ in eq. (\ref{21s}), remaining
smaller everywhere else. In fact, the relation $\overline{I}<S$ is valid for
the Gaussian ensembles of an arbitrary dimension $N$. As we have already
mentioned, for $N\gg 1$,
correlational entropy is equal to $S=\ln N$ whereas mean information entropy
reaches in this limit the value $\overline{I}=\ln (0.48 N)$.

At the same time, it can be shown by direct calculation that correlational
entropy remains at the maximum value also when variances of all matrix
elements have arbitrary values and the vector ${\bf n}$ ceases to be fully
isotropic. Indeed, the probability distributions for ensembles of real
symmetric matrices,
\begin{equation}
{\cal P}_{sym}({\bf n})d\varphi=\frac{\sigma\varsigma}
{\varsigma^2\cos^2\varphi+\sigma^2\sin^2\varphi}\,\frac{d\varphi}{2\pi},
                                                            \label{PGOE}
\end{equation}
and hermitian matrices,
\begin{equation}
{\cal P}_{herm}({\bf n})d\Omega=\frac{\sigma^2\varsigma_1^2\varsigma_2^2}
{\left[\varsigma_1^2\varsigma_2^2\cos^2\theta+
\sigma^2\left(\varsigma_1^2\sin^2\varphi+\varsigma_2^2\cos^2\varphi\right)
\sin^2\theta\right]^{3/2}}\,\frac{d\Omega}{4\pi},            \label{PGUE}
\end{equation}
remain even which yields $\overline {\bf n}=0$ as before. In
eq. (\ref{PGUE}), $\sigma^2=\left(\sigma_1^2+\sigma_2^2\right)/2$ whereas
$\sigma_{1,2}$ and $\varsigma_{1,2}$ are variances of diagonal and
off-diagonal matrix elements, respectively.

Correlational entropy drops to zero only when the off-diagonal matrix element
vanishes identically.  The randomness of interaction is of major importance.
On the contrary, information entropy is sensitive to the ratios of variances
which depend on the representation basis.

\section{Harmonic oscillator in a random field}

Here we consider a problem with infinite Hilbert space, namely that of a
linear harmonic oscillator in a uniform external field of a random strength.
The diagonalization of the hamiltonian at any fixed parameter value is
elementary but the diagonalization of the density matrix and calculation of
correlational entropy are not trivial and we did not encounter the full
solution in the literature although our results slightly overlap with those
of ref. \cite{osc}.

\subsection{The model}

At a given strength $F$, the hamiltonian of the problem,
\begin{equation}
H\equiv H_{0}-Fx=\frac{1}{2m}p^{2}+\frac{1}{2}m\omega^{2}\left(x-
\frac{F}{m\omega^2}\right)^{2} -\frac{F^2}{2m\omega^2},   \label{25}
\end{equation}
describes the harmonic oscillator with the shifted equilibrium point,
unchanged frequency and the energy spectrum
\begin{equation}
E_{n}(\lambda)=\hbar\omega(n+1/2-\lambda^{2})                   \label{26}
\end{equation}
where the integer $n$ plays the role of the exact quantum numbers $\alpha$ in
(\ref{1}), and the dimensionless ``noise" parameter
\begin{equation}
\lambda= \frac{F}{(2\hbar m\omega^{3})^{1/2}}               \label{27}
\end{equation}
displaces the spectrum as a whole (no crossing in this model).

The ``natural" basis $|k\rangle$ of equidistant non-shifted states
of a noiseless system, $\lambda=0$, is built
with the annihilation and creation operators,
\begin{equation}
x=\frac{x_{0}}{\sqrt{2}}(a^{\dagger}+a), \quad p=\frac{i\hbar}{\sqrt{2}x_{0}}
(a^{\dagger}-a), \quad x_{0}=\sqrt{\frac{\hbar}{m\omega}}.     \label{28}
\end{equation}
The shift of the equilibrium along the $x$-axis to the point
$\overline{x}=F/m\omega^{2}
=\sqrt{2}x_{0}\lambda$ is the unitary transformation
\begin{equation}
D(\lambda)=e^{-i\sqrt{2}\lambda x_{0}p/\hbar}=e^{\lambda(a^{\dagger}-a)}
=D^{\dagger}(-\lambda)                           \label{29}
\end{equation}
with the obvious properties
\begin{equation}
D(\lambda)aD(-\lambda)=a-\lambda, \quad D(\lambda)a^{\dagger}D(-\lambda)
=a^{\dagger}-\lambda                                           \label{30}
\end{equation}
so that (compare (\ref{26}))
\begin{equation}
H(\lambda)=D(\lambda)H_{0}D(-\lambda)-\lambda^{2}\hbar\omega.   \label{31}
\end{equation}
The eigenvectors $|n;\lambda\rangle$ at a given $\lambda$ are obtained from
the original non-shifted states $|n\rangle\equiv|n;\lambda=0\rangle$ by
the shift operator $D(\lambda)$,
\begin{equation}
|n;\lambda\rangle=D(\lambda)|n\rangle.                         \label{32}
\end{equation}
Therefore the amplitudes of the eigenvectors (\ref{1}) with
respect to the noiseless basis $|k\rangle$ are given by the matrix elements
\begin{equation}
C^{n}_{k}(\lambda)=\langle k|D(\lambda)|n\rangle                \label{33}
\end{equation}
of the shift operator (\ref{29}).
The instantaneous eigenvector $|n;\lambda\rangle$ carries an average number
of the original quanta equal to
\begin{equation}
\bar{n}=\langle n;\lambda|a^{\dagger}a|n;\lambda\rangle=n+\lambda^{2}.
                                                                \label{an}
\end{equation}

\subsection{Integral equation for the density matrix}

For the noise ensemble characterized by the (positively defined)
distribution function ${\cal P}(\lambda)$, the density matrix (\ref{5})
of the energy term $|n\rangle$ is
\begin{equation}
\rho^{(n)}_{kk'}=\int d\lambda {\cal P}(\lambda)\langle k|D(\lambda)|n\rangle
\langle n|D^{\dagger}(\lambda)|k'\rangle                          \label{34}
\end{equation}
where the unitarity of the $D$-transformation was used. This matrix
represents the density operator
\begin{equation}
\rho^{(n)}=\int d\lambda {\cal P}(\lambda)\,D(\lambda)|n\rangle
\langle n|D^{\dagger}(\lambda)                                 \label{34op}
\end{equation}
in the noiseless basis. It is convenient to project instead the eigenvalue
problem
\begin{equation}
\rho^{(n)}|\nu)=\rho^{(n)}_{\nu}|\nu)       \label{35}
\end{equation}
onto the set of states $|n;\lambda\rangle$ with the noise $\lambda$ treated
as the representation variable. Generally speaking, this set of states is
not complete and maps the original problem on the subspace of eigenstates
which belong to all nonvanishing eigenvalues $\rho^{(n)}_{\nu}$.
In the new representation, the eigenfunctions
\begin{equation}
\phi^{(n)}_{\nu}(\lambda)=A^{(n)}_{\nu}\sqrt{{\cal P}(\lambda)}\langle n|
D^{\dagger}(\lambda)|\nu),                                     \label{36}
\end{equation}
where $A_{\nu}^{(n)}$ are normalization constants, satisfy the integral
equation
\begin{equation}
\int d\lambda'\rho^{(n)}(\lambda,\lambda')\phi^{(n)}_{\nu}(\lambda')=
\rho_{\nu}^{(n)}\phi^{(n)}_{\nu}(\lambda)                        \label{37}
\end{equation}
with the real symmetric kernel
\begin{eqnarray}
\rho^{(n)}(\lambda,\lambda')=\sqrt{{\cal P}(\lambda)}
\langle n|D(\lambda'-\lambda)|n \rangle\sqrt{{\cal P}(\lambda')}. \label{38}
\end{eqnarray}
The diagonal elements of this kernel are simply the probabilities of the noise
distribution,
\begin{equation}
\rho^{(n)}(\lambda,\lambda)={\cal P}(\lambda).                    \label{39}
\end{equation}
They describe the ensemble itself and do not depend on the term $n$
under consideration.

The kernel (\ref{38}) can be expressed as
\begin{equation}
\rho^{(n)}(\lambda,\lambda')=\sqrt{{\cal P}(\lambda){\cal P}(\lambda')}
e^{-(\lambda-\lambda')^{2}/2}L_{n}[(\lambda-\lambda')^{2}].       \label{43}
\end{equation}
in terms of the Laguerre polynomial defined by the series
\begin{equation}
L_{n}(\xi)=\sum_{m=0}^{n}\frac{n!}{(m!)^2(n-m)!}\xi^{2m}. \label{40}
\end{equation}
It follows from (\ref{34op}-\ref{36}) that the orthonormalized,
$(n;\nu'|n;\nu)=\delta_{\nu'\nu}$, subset of eigenvectors belonging to nonzero
eigenvalues $\rho^{(n)}_{\nu}$ is expressed in terms of the functions
(\ref{36}) as
\begin{equation}
|n;\nu)=\left(\rho^{(n)}_{\nu}\right)^{-1/2}\int d\lambda
\sqrt{{\cal P}(\lambda)}\phi^{(n)}_{\nu}(\lambda)|n;\lambda\rangle  \label{ev}
\end{equation}
if the mutually orthogonal solutions (\ref{36}) of the integral equation
(\ref{37}) with the hermitian kernel (\ref{38}) are normalized to unity
which corresponds to the choice of
$A^{(n)}_{\nu}=\left(\rho^{(n)}_{\nu}\right)^{-1/2}$.

Starting with the
completeness condition $\sum_{\nu}|n;\nu)(n;\nu|=1$, one easily comes to the
kernel representation
\begin{equation}
\rho^{(n)}(\lambda,\lambda')=\sum_{\nu}\rho^{(n)}_{\nu}
\phi^{(n)}_{\nu}(\lambda)\phi^{(n)}_{\nu}(\lambda').        \label{ev1}
\end{equation}
In the original oscillator basis $|k\rangle\equiv |k;0\rangle$, the components
of the eigenvectors (\ref{36}) are
\begin{equation}
\left(\psi^{(n)}_{\nu}\right)_{k}
=\langle k|n;\nu)=\left(\rho^{(n)}_{\nu}\right)^{-1/2}
\int d\lambda\sqrt{{\cal P}(\lambda)}\phi^{(n)}_{\nu}(\lambda)
\langle k|D(\lambda)|n\rangle.                         \label{evk}
\end{equation}
Matrix elements of the shift operator are equal to
\begin{equation}
\langle k|D(\lambda)|n\rangle=e^{-\lambda^{2}/2}\sqrt{\frac{n!}{k!}}
(-\lambda)^{k-n}L_{n}^{k-n}(\lambda^{2})                    \label{42}
\end{equation}
where
\begin{equation}
L_{n}^{q}(\xi)=\sum_{m=0}^{n}\frac{(n+q)!}{m!(m+q)!(n-m)!}\xi^{2m} \label{40a}
\end{equation}
is the associated Laguerre polynomial, $L_{n}^{0}(\xi)\equiv L_{n}(\xi)$.
Due to the symmetry property
\begin{equation}
L_{n}^{k-n}(\xi)=\frac{k!}{n!}(-\xi)^{n-k}L_{k}^{n-k}(\xi),     \label{41}
\end{equation}
the matrix (\ref{42}) is symmetric with respect to $k$ and $n$.

Up to this point, the results are valid for any energy term
$|n;\lambda\rangle$ and an arbitrary distribution function ${\cal
P}(\lambda)$.

\subsection{Gaussian noise}

As an example of practical importance, we show here the exact solution for
an oscillator originally in the ground state, $n=0$, interacting with a
source of Gaussian noise of a certain width $\Lambda$.
The distribution function of the noise intensity is in this case
\begin{equation}
{\cal P}(\lambda)=\frac{1}{\sqrt{2\pi\Lambda^{2}}}
e^{-\lambda^{2}/2\Lambda^{2}},                            \label{44}
\end{equation}
and the ground state density matrix (\ref{43}) reads in the
$\lambda$-representation (we omit the superscript $n=0$)
\begin{equation}
\rho(\lambda,\lambda')=\frac{1}{\sqrt{2\pi\Lambda^2}}
e^{-(\lambda^{2}+\lambda'^{2})/4\Lambda^{2}}
e^{-(\lambda-\lambda')^{2}/2},                          \label{46}
\end{equation}
Using the properties of the Hermite polynomials
${\sl H}_{\nu}(x)$, one can represent the kernel (\ref{46}) as
\begin{equation}
\rho(\lambda,\lambda')=\sum_{\nu=0}^{\infty}\rho_{\nu}\phi_{\nu}(\lambda)
\phi_{\nu}(\lambda')                                  \label{ker}
\end{equation}
in terms of the set of orthonormalized functions
\begin{equation}
\phi_{\nu}(\lambda)=\left(\frac{\sinh\eta}{\pi}\right)^{\frac{1}{4}}
\frac{1}{\sqrt{2^{\nu}\nu!}}e^{-\lambda^2\sinh\eta/2}
{\sl H}_{\nu}(\sqrt{\sinh\eta}\lambda),                     \label{47}
\end{equation}
where
\begin{equation}
\rho_{\nu}=2\sinh(\eta/2)e^{-(\nu+1/2)}, \label{49}
\end{equation}
and the the parameter $\eta$, $0\leq \eta<\infty$, is defined by
\begin{equation}
\sinh\eta=\frac{\sqrt{1+4\Lambda^{2}}}{2\Lambda^{2}}.           \label{50}
\end{equation}

The representation (\ref{ker}) obviously solves the eigenvalue problem
(\ref{37}) for the ground state term.
The normalization of the density matrix $\sum_{\nu=0}^{\infty}\rho_{\nu}= 1$
can be checked directly.  The value of correlational entropy
defined by this density matrix is
\begin{equation}
S=-\sum_{\nu}\rho_{\nu}\ln\rho_{\nu}=\frac{\eta}{2}\coth(\eta/2)-
\ln[2\sinh(\eta/2)].                                            \label{51}
\end{equation}
The distribution function for the number of original quanta is given by the
diagonal matrix elements of the density matrix in the unperturbed basis
$|k\rangle$,
\begin{equation}
f_{k}=\langle k|\rho|k\rangle=\int d\lambda {\cal P}(\lambda)\left |\langle k
|0;\lambda\rangle\right|^{2}=\sqrt{\frac{2}{\pi}}\,\frac{\Gamma(k+1/2)}
{\Gamma(k+1)}\frac{\sinh(\eta/2)}{(\cosh\eta)^{k+1/2}}.       \label{51a}
\end{equation}
The average number of excited quanta equals
\begin{equation}
\bar{k}=\sum_{k=0}^{\infty}kf_{k}=\int d\lambda \lambda^{2}{\cal P}(\lambda)=
\overline{\lambda^{2}}=\Lambda^{2},                     \label{51b}
\end{equation}
which leads to the simple value of average energy of the term
\begin{equation}
\overline{E}={\rm Tr}(\rho H)=\hbar\omega(\bar{k}+1/2)-\overline{Fx}=
\hbar\omega(1/2-\Lambda^{2}),                                  \label{51c}
\end{equation}
in agreement with (\ref{26}) and (\ref{an}) for $n=0$.
The average energy of a given term
is always lowered by the presence of noise; however, this energy cannot be
attributed to the oscillator itself because it includes the contribution of
the external noise source.

According to (\ref{51b}) and (\ref{50}), the mean number of original quanta
excited by the external field can be written as
\begin{equation}
\bar{k}=\frac{1}{4\sinh^{2}(\eta/2)}.                       \label{51d}
\end{equation}
Introducing $\bar{k}$ as a new parameter,
\begin{equation}
\sinh\frac{\eta}{2}=1/(2\sqrt{\bar{k}}), \quad
\cosh\eta=1+1/(2\bar{k}),                             \label{51e}
\end{equation}
we obtain the distribution of quanta (\ref{51a}) in the form
\begin{equation}
f_{k}=(2\bar{k})^{-1/2}\,\frac{(2k)!}{2^{2k}(k!)^{2}}\,
\frac{1}{[1+1/(2\bar{k})]^{k+1/2}}.                             \label{51f}
\end{equation}
When the external perturbation is turned off, $\bar{k}\rightarrow 0$ and
$f_{k}\rightarrow \delta_{k0}$. In the opposite limit of strong noise, the mean
number of excited quanta is large, $\bar{k}\gg 1$, and eq.
(\ref{51f}) reduces to the chi-square (Porter-Thomas) distribution,
\begin{equation}
f_{k}=(2\pi\bar{k}k)^{-1/2}\,\exp[-k/(2\bar{k})], \quad \bar{k}\gg 1.
                                                                  \label{51g}
\end{equation}

We have to mention that, in the case of a fixed perturbation of a strength
$\lambda$, we would have a pure coherent state of a shifted oscillator with the
Poisson distribution
\begin{equation}
f_{k}=e^{-\bar{k}}\frac{\bar{k}^{k}}{k!},                    \label{51h}
\end{equation}
which reduces to the Gaussian distribution near the center $k=\bar{k}$ in
the classical limit of $\bar{k}\gg 1$.  The width of the coherent state in
the occupation number representation is $(\Delta k)^{2}=\bar{k}$ and
therefore the relative uncertainty of this number is small, $\Delta
k/\bar{k}=\bar{k}^{-1/2}$.  Contrary to that, in our random noise ensemble
(\ref{51g}) the state is mixed and the fluctuations of the number of quanta
are not suppressed even in the classsical limit,
\begin{equation}
\overline{(\Delta k)^{2}}=2\bar{k}^{2}.                         \label{51i}
\end{equation}
The situation is similar to that in the boson intensity interferometry (Hanbury
Brown - Twiss) for many incoherent sources, see for example \cite{And}.

\subsection{Relation to thermodynamics}

The eigenfunctions (\ref{47}) of the density matrix (\ref{46})
in the $\lambda$-representation formally coincide with the coordinate wave
functions $(\xi|\nu)$ for
a linear oscillator with the dimensionless ``coordinate"
$\xi\equiv\sqrt{\sinh\eta}\lambda$. The dimensionless
hamiltonian for such an oscillator would be
\begin{equation}
{\cal H} = \frac{1}{2}\left(\xi^2-d^2/d\xi^2\right) =
 (\alpha^{\dag}\alpha+1/2)                                     \label{52}
\end{equation}
where the annihilation and creation operators
\begin{equation}\label{53}
\alpha=\frac{1}{\sqrt{2}}(\xi+d/d\xi),\quad
\alpha^{\dag}=\frac{1}{\sqrt{2}}(\xi-d/d\xi)
\end{equation}
are introduced. The energy spectrum of the operator (\ref{52}),
$\varepsilon_{\nu}=(\nu+1/2)$, is equidistant and quantized in
$\lambda$-independent units at any noise magnitude. Since the density
operator is diagonal in the basis of the eigenfunctions of (\ref{52}), it
is tempting to interpret our results as an equilibrium reached by the system
under the influence of the noise. The equilibrated (``dressed") system is
represented by the effective oscillator (\ref{52}). We can assign the
physical frequency $\omega$ to this oscillator and interpret the
corresponding occupation number spectrum (\ref{49}) as the Planck formula
with the effective temperature
\begin{equation}
T=\frac{\hbar\omega}{\eta}.                            \label{54}
\end{equation}
Under this identification, the temperature scale is related to the range of
noise,
\begin{equation}
\Lambda=\frac{1}{2\sinh(\hbar\omega/2T)},               \label{55}
\end{equation}
or to the mean square value of the random force,
\begin{equation}
(\overline{F^{2}})^{1/2}x_{0}=\frac{1}{\sqrt{2}}\,\frac{\hbar\omega}
{\sinh(\hbar\omega/2T)}.                                   \label{56}
\end{equation}

The density operator can be represented in terms of the effective
oscillator defined above as
\begin{equation}                                              \label{57}
\rho = \exp\left(-\hbar\omega{\cal H}/T\right) =
\sum\limits_{\nu=0}^{\infty}\rho_{\nu}|\nu)(\nu|,\quad
\rho_{\nu} = 2\sinh\left(\frac{\hbar\omega}{2T}\right)
\exp\left(-\frac{\hbar\omega}{T}(\nu+1/2)\right).
\end{equation}
The effective oscillator is in the conventional thermodynamic equilibrium
with a heat bath of temperature $T$, eq. (\ref{54}). In an ordinary
way, we can define its free energy
\begin{equation}                                                \label{58}
{\cal F} = T\ln\left[2\sinh\left(\frac{\hbar\omega}{2T}\right)\right]
\end{equation}
so that canonical thermodynamic entropy
\begin{equation}                                                  \label{59}
S = -\frac{\partial {\cal F}}{\partial T}.
\end{equation}
coincides with correlational entropy (\ref{51}).
Thermodynamic energy of the effective oscillator is equal to
\begin{equation}                                                  \label{60}
{\cal U} = {\cal F}+TS = \hbar\omega\left(\bar{\nu}+\frac{1}{2}\right)=
\frac{\hbar\omega}{2}\coth\frac{\hbar\omega}{2T}
\end{equation}
where the mean number of excited effective quanta is given by the Planck
formula,
\begin{equation}
\bar{\nu}=\frac{1}{\exp(\hbar\omega/T)-1}.                    \label{60a}
\end{equation}

Let us consider an arbitrary dynamical variable of the original oscillator
described by the operator
${\cal O}(a^{\dag},a)$,
\begin{equation}                                                \label{61}
\overline{{\cal O}} = \int d\lambda {\cal P}(\lambda)
\langle\lambda|{\cal O}(a^{\dag}, a))|\lambda\rangle =
\int d\lambda {\cal P}(\lambda) O(\lambda,\lambda),
\end{equation}
where the operators $a^{\dag}$ and $a$ of original quanta are
supposed to be normally ordered with respect to the unperturbed vacuum;
in the second equality (\ref{61}) the properties of the coherent state
$|\lambda\rangle$ were used.  Since the parameters of the effective
oscillator depend on temperature (\ref{54}) via the factor
$\sqrt{\sinh\eta}$, the mean value of any operator (\ref{61}) contains an
additional nontrivial temperature dependence. Indeed, using the relations
(\ref{39}) and (\ref{ker}),
\begin{equation}                                               \label{62}
{\cal P}(\lambda) = \rho(\lambda,\lambda) = \sum\limits_{\nu=0}^{\infty}
\rho_{\nu}\phi_{\nu}(\lambda)\phi_{\nu}(\lambda),
\end{equation}
we can express the mean value (\ref{61}) in terms of the effective
oscillator
$$\overline{O} = \sum\limits_{\nu=0}^{\infty}\rho_{\nu}
(\nu|O\left(\frac{\alpha^{\dag}+\alpha}{\sqrt{2\sinh\eta}},
\frac{\alpha^{\dag}+\alpha}{\sqrt{2\sinh\eta}}\right)|\nu) =$$
\begin{equation}                                                  \label{63}
{\rm Tr}\left[\rho\,O\left(\frac{\alpha^{\dag}+\alpha}
{\sqrt{2\sinh\left(\hbar\omega/T\right)}},\frac{\alpha^{\dag}+
\alpha}{\sqrt{2\sinh\left(\hbar\omega/T\right)}}\right)\right]
\end{equation}
with the density operator from eq. (\ref{57}). For example, the mean
number of original quanta is given by
\begin{equation}                                                 \label{64}
\bar{k} = \sum\limits_{\nu=0}^{\infty}\rho_{\nu}
(\nu|\left(\frac{\alpha^{\dag}+\alpha}{\sqrt{2\sinh\eta}}\right)^2|\nu) =
\frac{\bar{\nu}+1/2}{\sinh\eta}=\frac{{\cal U}}{\hbar\omega\sinh\eta} =
\frac{1}{4\sinh^2(\hbar\omega/2T)}
\end{equation}
in agreement with (\ref{51d}). The distribution of the original quanta
differs from the Planck formula (\ref{60a}).  At a narrow range of noise,
$T\ll \hbar\omega,\; \Lambda\approx \exp(-\hbar\omega/2T)$, which
corresponds to the quantum low-temperature limit, we obtain a normal quantum
result,
\begin{equation}                                            \label{65}
\bar{k}= e^{-\hbar\omega/T}, \quad T\ll \hbar\omega.
\end{equation}
A broad noise, $\Lambda\approx T/\hbar\omega\gg 1$, leads to the classical
limit of high temperature. In this limit, the number of quanta is increasing
with temperature quadratically rather than linearly,
\begin{equation}                                               \label{66}
\bar{k}=\left( \frac{T}{\hbar\omega} \right)^{2}, \quad T\gg \hbar\omega.
\end{equation}

Let the random field applied to an oscillator with the electric charge $e$
be abruptly removed at the moment $t=0$. The oscillator remains excited and
starts to radiate electromagnatic waves losing its energy with the rate
\begin{equation}                                             \label{67}
\overline{\dot{E}} = \frac{2e^2 x_0^2\omega^4}{3c^3}\bar{k} =
\frac{2e^2 x_0^2\omega^4}{3c^3}
\frac{\bar{\nu}+1/2}{\sinh(\hbar\omega/T)} =
\frac{2e^2 x_0^2\omega^4}{3c^3}\frac{1}{4\sinh^2(\hbar\omega/2T)}.
\end{equation}
The emission rate directly measures {\sl effective} temperature. Having no
knowledge about the noise properties and detecting only the frequency
$\omega$ and the intensity $\overline{\dot{E}}$ of radiation, we could assume
thermal equilibrium inside the radiating system and assign temperature
$T_{0}$ by the direct application of the Planck formula instead of the actual
distribution (\ref{51f}). It is easy to see that these temperatures are
related through
\begin{equation}
\sinh^{2}(\hbar\omega/2T)=\frac{1}{2}\sinh(\hbar\omega/2T_{0})
e^{\hbar\omega/2T_{0}}.
                                                             \label{68}
\end{equation}
For the narrow noise, two definitions agree,
\begin{equation}
T_{0}\approx T, \quad T\ll \hbar\omega.                      \label{69}
\end{equation}
However, under conditions of the broad noise, they lead to very different
assignments with $T_{0}\gg T$,
\begin{equation}
T_0\approx \frac{T^{2}}{\hbar\omega}, \quad T\gg \hbar\omega.   \label{70}
\end{equation}
Two definitions can be discriminated in the case of the spectrum of normal
modes with different frequencies by the deviations from the Planck formula
at the low-frequency edge, if the noise amplitude $F$ is not frequency
dependent.

\section{A many-body example}

Here we illustrate the concept of correlational entropy by calculating this
quantity for a realistic many-fermion model with strong interparticle
interaction of a
variable random strength. We use for this purpose the version of
the nuclear shell model which provides the best available description of
spectroscopic data for all $sd$-nuclei \cite{WB}. The model hamiltonian
consists of the one-body part given by a spherical mean field due to the closed
shell core of $^{16}$O and the semiempirical residual two-body interaction
determined by 63 reduced matrix elements
$\langle(j_{1}j_{2})JT||V||(j_{3}j_{4})JT\rangle$ allowed in this truncated
Hilbert space by the conservation laws of angular momentum $J$ and isospin $T$.
Apart from numerous specific nuclear calculations, this model was recently used
\cite{entr,temp,big} for studying highly excited states beyond the limits
of experimental information. One of the methods utilized in this analysis
studies the evolution of observables as a function of the strength of the
residual interaction \cite{big,kus1,fraz}.

We take as a generic example a set of states with quantum numbers $J^{\pi}T=
0^{+}0$ in the $^{24}$Mg nucleus with 4 protons and 4 neutrons in the
$sd$-shell model space. The dimension of this set of states is $N=$325. The
unperturbed basis $|k\rangle$ is that of simple independent particle
configurations which are projected onto correct values of integrals of motion
$J$ and $T$. The residual interaction is split into two parts, diagonal and
off-diagonal with respect to the basis $|k\rangle$. The diagonal part is
included into unperturbed energies in order to lift the degeneracy of bare
configurations. The off-diagonal interaction with the overall factor $\lambda$
in front is our ``random noise" with $\lambda=0$ and $\lambda=1$ corresponding
to the independent particle case and to the realistic strength, respectively.

The many-body hamiltonian is diagonalized as a function of $\lambda$ which
provides us with 325 energy terms $E_{\alpha}(\lambda)$. These terms were shown
in Fig. 1 of Ref. \cite{kus1}. Multiple avoided level crossings rapidly
transform the eigenstates of the hamiltonian into complicated superpositions of
many original configurations. The generic signatures of quantum chaos in level
statistics, such as nearest level spacing distribution, spectral rigidity, and
level curvature distribution, agree with the GOE predictions already at
$\lambda\approx 0.3$. The calculation of information entropy in the original
mean field basis shows that the evolution of complexity continues so that at
$\lambda=1$ the states in the middle of the spectrum are close to the GOE limit
of $I=\ln(0.48 N)$.

Using the known eigenfunctions $|\alpha\rangle$
of the system in the original basis $|k\rangle$, we find the density matrix
$\rho^{(\alpha)}_{kk'}(\lambda)$ of Eq. (\ref{2})
and construct the ensemble of such matrices for
different values of $\lambda$. The uniform averaging over $\lambda$ in the
chaotic interval $0.3 \leq \lambda \leq 1$ gives the density matrix
$\rho^{(\alpha)}$ of Eq. (\ref{5}). This matrix can be diagonalized and its
eigenvalues (occupation numbers) determine correlational entropy (\ref{7}).

Fig. 3 shows correlational entropy $S^{\alpha}$ of all 325 states
ordered according to increasing energy. Although the
fluctuations at this dimension are significant, nevertheless we see the same
qualitative pattern as the one obtained with the help of information entropy
\cite{entr,big}. The degree of mixing measured by correlational entropy evolves
along the spectrum revealing the regular increase with excitation energy (in
finite Hilbert space the statistical properties of eigenstates are symmetric
with respect to the middle of the spectrum; in the thermal description
\cite{temp} the center corresponds to infinite temperature with positive and
negative temperature regions on the left and on the right, respectively). The
behavior of information entropy and of single-particle entropy defined
by the evolution of the single-particle occupation numbers is essentially the
same. However, the absolute scales are very different. As seen in Fig. 3,
the typical values of $\tilde{N}_{\alpha}=\exp[S^{(\alpha)}]$ for the most
complicated states are close to 6 which is much lower than the localization
length $\exp[I^{(\alpha)}]$ found with the use of information entropy.
From the perturbative analysis of Sect. 3 we know that the number of
effectively occupied eigenstates of the density matrix is related to the order
of perturbation theory. Then $\tilde{N}$ can be interpreted in terms
of a number of particle-hole excitons created by the spectral evolution.
This point deserves to be studied more in detail.

\section{Conclusion}

We have discussed some examples of quantum-mechanical systems under the
conditions of stationary external noise which was defined by a distribution
function of parameters controlling the interaction hamiltonian.  The maximum
of quantum information on such systems is provided by the density matrix
which was found for various cases, analytically in Sect. 3-5 and numerically
in Sect. 6.  The perturbative treatment of Sect. 3 illustrates the mechanism
of equlibration through the consecutive selection of directions in Hilbert
space which form the eigenbasis of the density matrix.  In other examples
the weakness of the interaction was not assumed, and the solution was exact.

We have used correlational (von Neumann) entropy as a tool for measuring the
degree of disorder and complexity of stationary states. By definition, this
entropy vanishes for pure quantum states. Therefore it characterizes the
system in a given noise ensemble when the stationary states are mixed. In
general, the arising configuration with certain occupation numbers of the
eigenstates of the density operator does not coincide with thermal
equilibrium as defined by canonical Gibbs ensembles. Accordingly,
correlational entropy is not equal to thermodynamic entropy. Moreover,
correlational entropy is calculated for the individual energy terms which
evolve adiabatically as a function of the noise strength. The resulting
state is in general different for different terms.

It would be interesting to determine the conditions for the external noise
which would give the same equilibrated state as in the heat bath. It is
assumed usually that the necessary ingredient is the continuous spectrum of
the normal modes represented in the spectral expansion of noise \cite{irr}.
We deal with the stationary noise represented by random parameters in the
hamiltonian. In Sect. 5 we have however shown that the ground state of a
harmonic oscillator in a random uniform field can be described as an
equilibrated thermal state of an effective oscillator with temperature
determined by the Gaussian width of the field distribution. The situation
here is similar to that in many-body physics where the interacting system
can be modeled \cite{qp} by a gas of dressed quasiparticles with properties
depending on temperature (in our case the energy spectrum of the oscillator
is not renormalized but the coordinate scale is determined by noise and
changes with effective temperature). The difference between a simply heated
oscillator in a thermostat and an oscillator excited by a Gaussian noise and
described with the aid of effective temperature might be important for the
problems of multiple meson production \cite{And,bambah} in high energy
collisions. We can remind also that the notion of the effective quantum
oscillator appears naturally in the problem of a uniformly accelerated
observer in the Minkowski world \cite{dav,bishop}.  An observer falling with
the proper acceleration $g$ sees the Minkowski vacuum as a black body
emitter with the effective temperature $T=\hbar g/(2\pi c)$. This
consideration is closely related to the Hawking black hole radiation
\cite{sciama}. However, in those cases, one has squeezed rather than
coherent states of the oscillator (in terms of the unperturbed system they
are produced by the source creating the quanta pairwise which is described
with the help of the Bogoliubov transformation).

Representing the complexity of individual quantum terms, our invariant
correlational entropy can be juxtaposed to representation-dependent
information entropy. We drew the attention of the reader to their similarity
and distinction in various applications. Although they are formally quite
different and may react in a different way on the change of parameters of a
simple regular system, Sect. 4, they behave qualitatively similar in a
complicated system, Sect. 6, where one sees the standard signatures of
quantum chaos. Due to similarity of adjacent states in the chaotic regime
\cite{big}, both entropies are smooth functions of excitation energy and
therefore can be considered as thermodynamic variables. Both entropies carry
information on complexity of individual states and its evolution along the
spectrum. Quantitatively, information entropy (in the appropriate basis!)
expresses this complexity in terms of a number of mixed simple
configurations whereas correlational entropy measures essentially similar
properties in larger blocks as a number of classes of states effectively
mixed by external noise. Of course, one should remember that information
entropy refers to a given hamiltonian while correlational entropy describes a
``system plus noise" complex. The further studies will bring the more deep
insight into the problem.\\
\\
\\
The authors acknowledge support from the NSF grants 94-03666, 95-12831 and
96-05207. One of us (V.V.S) thanks National Superconducting Cyclotron
Laboratory for hospitality; he acknowledges financial support from INTAS,
grant 94-2058. V.V.S. and V.Z. thank V.F. Dmitriev and V.B.
Telitsin for interesting and useful discussions; they are also grateful to
D.V. Savin and V.V. Zelevinsky for assistance.

\newpage

{\bf Figure captions}\\[0.2cm]
\noindent
{\bf Figure 1}. Correlational entropy of the two-level system (16)
as a function of the ratio
$\tau=2V/\Lambda$ of the strength of the mixing interaction $2V$ to the
range of the random fluctuation $\Lambda$
of the level positions; parts {\sl a,b,c} and {\sl d}
correspond to the values $\chi=\epsilon/\Lambda=0.2,\;0.6,\;1$ and 5,
respectively, where $\epsilon$ is the static level spacing.\\
\\
{\bf Figure 2}. Correlational entropy of the two-level system (16) as a
function of the random perturbation strength $\chi^{-1}$ for four values
of the intrinsic mixing strength $\zeta=\tau/\chi$ : 0.01 (solid line),
0.1 (dots), 2 (dash), 10 (dot-dash). All the curves asymptotically approach
the limiting value $\ln2$; the rate of approach is slower for larger
values of the strength parameter.\\
\\
{\bf Figure 3}. Correlational entropy of 325 $J^{\pi}T=0^{+}0$ states in the
$sd$-shell model for the $^{24}$Mg nucleus. The states are ordered in
increasing energy.

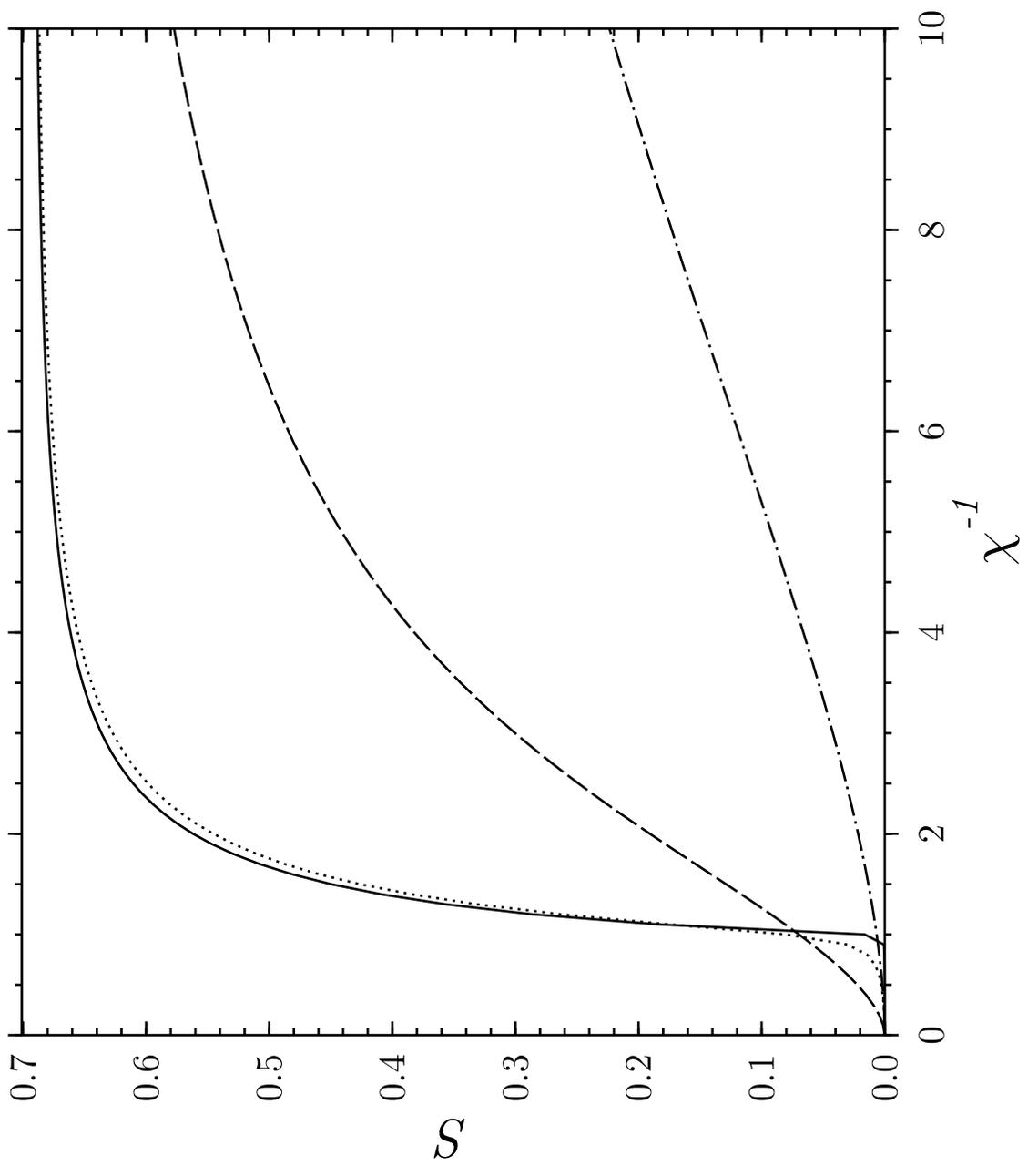
\end{document}